%% file: main.tex
\documentclass[journal]{IEEEtran}
\usepackage{amsmath,amsfonts}
\usepackage{algorithm}
\usepackage{array}
\usepackage{mathtools}
\usepackage{xcolor}
\usepackage[caption=false,font=normalsize,labelfont=sf,textfont=sf]{subfig}
\usepackage{textcomp}
\usepackage{stfloats}
\usepackage{url}
\usepackage{verbatim}
\usepackage{graphicx}
\usepackage{cite}
\usepackage{macro_expression}
\usepackage{algorithm}
\usepackage{algpseudocode}
\usepackage{algorithmicx}
\algrenewcommand\algorithmicrequire{\textbf{Input:}}
\algrenewcommand\algorithmicensure{\textbf{Output:}}

\usepackage{xcolor}
\begin{document}
% \title{Consensus-based Distributed Target Tracking \\ using FMCW Radar Networks}

\title{ Distributed Target Tracking using Radar Networks}
% \author{Shao-Hsuan Hung,~\IEEEmembership{,~IEEE,}
%         % <-this % stops a space
% \thanks{This paper was produced by the IEEE Publication Technology Group. They are in Piscataway, NJ.}% <-this % stops a space
% \thanks{Manuscript received April 19, 2021; revised August 16, 2021.}}
% \author{\IEEEauthorblockN{Shao-Hsuan Hung} \\
% \IEEEauthorblockA{\textit{Signal Processing Systems, EEMCS} \\
% \textit{Delft University of Technology}\\
% Delft, the Netherlands \\
% s.h.hung-1@tudelft.nl}
% \thanks{Version: \today}
% \thanks{This work is funded by the EU-HORIZON-KDT-JU-2023-2-RIA,
% under grant agreement No 101139996, the ShapeFuture project - ”Shaping
% the Future of EU Electronic Components and Systems for Automotive Applications”.}
% }

\author{
\IEEEauthorblockN{Shao-Hsuan Hung}\\
\IEEEauthorblockA{
\textit{Signal Processing Systems Group, Faculty of EEMCS}\\
\textit{Delft University of Technology}\\
\textit{Delft, The Netherlands}\\
Email:\texttt{shaohung@tudelft.nl}}

\and

\IEEEauthorblockN{Raj Thilak Rajan}\\
\IEEEauthorblockA{
\textit{Signal Processing Systems Group, Faculty of EEMCS}\\
\textit{Delft University of Technology}\\
\textit{Delft, The Netherlands}\\
Email: \texttt{r.t.rajan@tudelft.nl}}
\thanks{Version: \today}
\thanks{This work is funded by the EU-HORIZON-KDT-JU-2023-2-RIA,
under grant agreement No 101139996, the ShapeFuture project - ”Shaping
the Future of EU Electronic Components and Systems for Automotive Applications”.}
}

% The paper headers
\markboth{Journal of \LaTeX\ Class Files,~Vol.~14, No.~8, August~2021}%
{Shell \MakeLowercase{\textit{et al.}}: A Sample Article Using IEEEtran.cls for IEEE Journals}
\maketitle

\input{chapter/abstract}
\input{chapter/introduction}
\input{chapter/problem_formulation}
\input{chapter/data_model}
\input{chapter/fusion_algor}
% \input{chapter/simulation}
\input{chapter/simulation_polish}
\input{chapter/conclusion}
\input{chapter/appendix}

\bibliographystyle{IEEEtaes}
\bibliography{main}
\end{document}

%% file: chapter/abstract.tex
\begin{abstract}
Distributed target tracking is essential for scalable and robust sensing systems, as it enables multiple radar nodes to cooperatively estimate a target state without relying on a centralized fusion center. In this paper, we present a fully distributed framework for single-target tracking in frequency-modulated continuous-wave (FMCW) radar networks. Each monostatic radar node observes local range and Doppler measurements and exchanges information only with neighboring nodes. Two consensus optimization-based estimators are developed. First, a Distributed Maximum A Posteriori (D-MAP) estimator is formulated for batch-based tracking, where prior state information is incorporated and the resulting optimization problem is solved using consensus-based alternating direction method of multipliers (ADMM). Second, a Distributed Extended Kalman Filter (D-EKF) is proposed for recursive tracking, where each node performs local prediction and correction followed by consensus ADMM-based information exchange. We derive the posterior Cramér-Rao lower bound (PCRLB) as a theoretical performance benchmark. Our simulation results show that D-MAP improves the accuracy of the estimation over the distributed maximum-likelihood baseline. These results demonstrate that the proposed framework provides a scalable and robust alternative to centralized radar tracking, given only local inter-node communication.
% This paper proposes a consensus-based distributed framework for single-target tracking for FMCW radar networks. Each monostatic, single input single output radar sensor collects local range and Doppler measurements and cooperates only with neighboring nodes to estimate the target state over time. Two distributed estimation strategies are developed. 
% First, a decentralized maximum a posteriori (MAP) approach is formulated and solved using consensus alternating direction method of multiplier (ADMM) to exploit prior state information in batch-based tracking. Second, a distributed extended Kalman filter (EKF) is introduced for recursive tracking when target dynamics are available, where local prediction and correction are followed by inter-node consensus to obtain a network-wide estimate. Simulation results for a network of multiple radar nodes show that the proposed MAP method improves estimation accuracy over distributed maximum-likelihood estimation, especially under low-SNR conditions, while the distributed EKF achieves stable tracking performance with accuracy close to that of a centralized solution. These results demonstrate that the proposed framework provides a scalable and robust alternative to centralized radar tracking, while avoiding single points of failure and requiring only local communication among neighboring sensors.
\end{abstract}

%% file: chapter/introduction.tex
\section{Introduction}
%%%%% Logic: 
% Why target tracking is important, and application
% What sensors are used
% Why radar and why distributed?
% 
%%%%% TODO: Add reference.
% General intro to target tracking
Target tracking aims to estimate the time-varying state of a target, typically its position and velocity, and is essential in applications such as drone surveillance~\cite{dron_tracking}, space situational awareness~\cite{space_SW}, rover navigation~\cite{rover_tracking}, and automotive perception~\cite{car_radar_tracking}. Tracking systems employ sensing modalities including cameras, LiDAR, ultrasonic sensors, and radio-frequency systems, each offering different trade-offs in coverage, robustness, and cost~\cite{sensor_review}.
% There are many way for tracking, but why we need radar. What the radar network is important, possiblly for wide-range applications: automotive radar perception and RSU for reconginizing targets.
Among these sensing modalities, radar plays a particularly important role in safety-critical and adverse-environment applications. Unlike vision-based sensors, radar operates reliably in poor illumination and under challenging weather conditions such as fog, rain, or snow~\cite{radar_vs_cam}. Compared to LiDAR, radar typically offers longer detection ranges, lower susceptibility to atmospheric scattering, and a more favorable cost–performance ratio in large-scale deployments~\cite{compare_sensor}. In addition to alternative radar waveforms such as OFDM radar~\cite{OFDM} and UWB radar~\cite{UWB_radar}, Frequency-Modulated Continuous-Wave (FMCW) radar is chosen for its ability to simultaneously measure range and velocity with high precision using low power, cost effective hardware~\cite{FMCWradar}. This makes it uniquely suited for distributed networks where low-cost, high-resolution sensing at each node is essential for robust state estimation.

% Motivtate why multiple sensor, why distributed radar
A single radar may provide incomplete target information because of occlusions, limited field of view, and unfavorable aspect angles. Radar networks mitigate these limitations through spatial diversity, as multiple nodes observe the same target from different locations and viewing directions, providing complementary range and Doppler information for estimating the target position and velocity~\cite{distributed_radar_net,ignore_clutter2}.
However, the use of multiple radars does not necessarily imply distributed inference. In centralized fusion, measurements or detections from all radar nodes are transmitted to a fusion center, where the final estimate is computed. Although this architecture can provide strong estimation performance, it remains vulnerable to single-point failures and may require substantial communication bandwidth as the number of nodes increases. A less communication-intensive alternative is decentralized, or track-level, fusion. This approach reduces the communication burden by allowing each node to transmit local estimates rather than raw measurements. However, the final inference still depends on a higher-level fusion structure and must account for cross-correlations among local tracks~\cite{ckc-decentralized_fusion,chen_track_fusion_limits,fusion_center1}. Fully distributed inference instead allows each node to update its estimate using local measurements and information exchanged only with neighboring nodes~\cite{fusion_arch_algor}. This architecture improves scalability, supports low-latency processing, and enhances robustness to node and communication-link failures. Throughout this paper, the term \emph{distributed fusion} refers to this fully distributed inference architecture.

% After motivation of distributed radar, we now go to the processing algorithm of distributed radar.
To bridge the gap between architectural potential and practical implementation, recent research has focused on distributed target tracking algorithms for distributed radar networks.
Jabbari \emph{et al.}~\cite{tar_loc_liter} developed a weighted least-squares estimator for instantaneous target-state estimation, while Chaganti \emph{et al.}~\cite{srikar} proposed a distributed maximum-likelihood estimator based on ADMM for range--Doppler localization without a fusion center. These methods do not explicitly exploit temporal dependencies between successive target states. Kim \emph{et al.}~\cite{kim2013multiple} incorporated target dynamics and data association within an extended Kalman filter, but retained a centralized processing architecture. Consequently, fully distributed radar tracking methods that explicitly model target dynamics and recursively update the target state remain limited.

% This reveals a significant gap between snapshot-based localization and practical multi-step tracking in distributed radar networks.
\begin{figure}[t]
    \centering
    \includegraphics[width=1\linewidth]{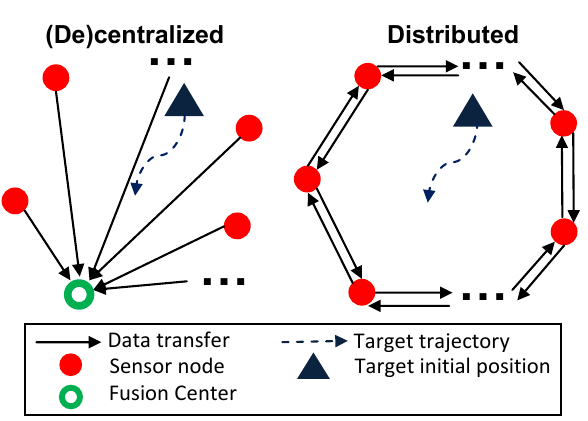}
    \caption{Illustration of centralized and distributed sensing frameworks. In the centralized framework (left), all sensor nodes transmit their measurements to a fusion center. In the distributed framework (right), each node exchanges information only with its neighboring nodes.}
    \label{fig:ctrl_vs_dectrl}
\end{figure}

In this paper, we extend our previous work~\cite{srikar} on target localization to target tracking using distributed radar networks. Each radar node processes its local range-Doppler measurements and exchanges information only with its neighboring nodes to cooperatively estimate the target state in a fully distributed manner, without the need for a fusion center, as illustrated in Fig.~\ref{fig:ctrl_vs_dectrl}. We consider both batch and recursive tracking: a distributed maximum a posteriori estimator solved using consensus ADMM and a distributed extended Kalman filter that also employs consensus ADMM for inter-node information fusion.

% Numerical results demonstrate that the proposed approach eliminates the single point of failure inherent in centralized architectures, scales with the network size, and attains tracking performance comparable to that of conventional centralized solutions while requiring only local inter-node communication.

The main contributions of this work on distributed target tracking using a decentralized radar network are summarized as follows:
\begin{itemize}
    \item We propose centralized fusion algorithms - centralized Maximum A Posteriori estimator (C-MAP), centralized Extended Kalman Filter (C-EKF) and corresponding distributed fusion algorithms - Distributed MAP (D-MAP), Distributed EKF (D-EKF) approach for single-target tracking, based on distributed optimization and local information exchange between neighboring nodes.
    \item We derive the posterior Cram\'er--Rao lower bound (PCRLB) as a theoretical performance benchmark for the proposed Bayesian tracking methods.
    \item We demonstrate through numerical simulations that D-MAP improves upon the distributed MLE baseline, while D-EKF achieves tracking performance close to its centralized estimators without relying on a fusion center.
\end{itemize}

\textbf{\textit{Outline:}} Section~\ref{sec: Problem statement} introduces the tracking problem and assumptions. Section~\ref{sec: data model} presents the radar signal and measurement models. Sections~\ref{sec: centralizede fusion} and~\ref{sec: distributed fusion} describe the centralized reference methods and the proposed distributed tracking algorithms, respectively. Section~\ref{sec: simulation} presents the numerical results and compares the proposed algorithms with the considered baselines. Finally, Section~\ref{sec: conclusion} summarizes our contributions and future direction.

\textbf{\textit{Notation:}} We use small and capital boldface letters to denote vectors and matrices respectively. $\mathbb{R}^N$ denotes the $N$-dimensional real vector space. For a matrix $\mathbf{A}$, $\mathbf{A}^\top$ denotes its transpose, $\mathbf{A}^{-1}$ its inverse (when it exists), and $\|\mathbf{A}\|$ its spectral norm. For a vector $\bsym{\theta}$, $\|\bsym{\theta}\|$ denotes its Euclidean norm. The notation $\mathbf{A}\succeq0$ means that the matrix $\mathbf{A}$ is positive semi-definite.  The notation $\bsym{\theta}\sim\mathcal{N}(\boldsymbol{\mu},\mathbf{\Sigma})$ denotes a Gaussian distribution with mean $\boldsymbol{\mu}$ and covariance $\mathbf{\Sigma}$. The operator $\mathrm{blkdiag}(\cdot)$ denotes a block-diagonal matrix. The weighted norm $\|\bsym{\theta}\|^2_\mathbf{M}$ is defined as $\bsym{\theta}^\top\mathbf{M}\bsym{\theta}$. The half vectorization of a symmetric matrix $\mathbf{A} \in \mathbb{R}^{n\times n}$ is denoted by $\operatorname{vec}_h(\mbf{A})\in \mathbb{R}^{n(n+1)/2}$, where elements are filled in column-major order vector, i.e., $\operatorname{vec}_h(\mbf{A})\coloneqq[\mbf{A}_{1,1},\cdots,\mbf{A}_{1n,},\mbf{A}_{2,2},\allowbreak \cdots,\mbf{A}_{2,n},\mbf{A}_{n-1,n-1},\allowbreak\cdots,\mbf{A}_{n-1,n},\mbf{A}_{n,n}]^\top$. Estimated quantities are denoted by a hat, and the subscripts $k|k-1$ and $k|k$ denote the prediction and correction steps, respectively. 
% Regarding indexing, $n$ is denoted as the sensor index, $k$ is utilized as the tracking time index, and $i$ represents the iteration index within optimization algorithm. -> This is skip, can be used for the later paragraph. 

%% file: chapter/problem_formulation.tex
\section{Problem Statement}
\label{sec: Problem statement}
% Few concrete setup should be mentioned here: (Aslo in the signal model part)
% Static monostatic radars

% About the network 
We consider a network of $N$ stationary omnidirectional monostatic, SISO FMCW radar nodes with known positions $\mathbf{g}_n = [x_n, y_n]^\top,\ \forall n = 1, 2,\cdots, N$, as shown in the left of Fig.\ref{fig:ctrl_vs_dectrl}. The network estimates position and velocity of a point target from noisy range and Doppler measurements. Communication among the radar nodes is represented by an undirected graph $\mathcal{G}=(\mathcal{V},\mathcal{N})$, where $\mathcal{V}$ and $\mathcal{N}$ denote the sets of radar nodes and communication links, respectively. The neighbor set of node $n$ is denoted by $\mathcal{N}_n$.

% About each node in the network
Each radar transmits a linear frequency-modulated signal with bandwidth $B$ and wavelength $\lambda$, consisting of $L$ pulses per coherent processing interval (CPI). One target-state estimate is obtained from $M$ CPIs, corresponding to an observation interval of $M t_{\mathrm{CPI}}$. For the batch-processing estimators (MLE and MAP), $K$ tracking updates therefore span a total observation duration of $T = K M t_{\mathrm{CPI}}$, with $k=1,\ldots,K$ denoting the tracking-time index. The indices $n$ and $i$ denote the radar node and algorithm iteration, respectively. In contrast, the recursive EKF-based estimator updates the target state sequentially as each new measurement becomes available.

To focus on the proposed distributed state-estimation framework, we adopt a detect-then-track (DTT) architecture and consider a high-SNR single-target scenario. Target detection and range–Doppler measurement extraction are assumed to be performed by the radar front end prior to tracking. The proposed estimator is therefore conditioned on successfully detected target-generated measurements, and clutter-induced false alarms and missed detections are not explicitly modeled. Standard radar detection and preprocessing techniques, such as clutter suppression and CFAR processing, can be employed to control false alarms and improve target detection performance in cluttered environments \cite{rohling1983cfar,shnidman1995detection,shnidman2005clutter}.

% To focus on the proposed distributed tracking framework, we consider a high-SNR single-target scenario and neglect clutter and missed detections, assuming that these effects are mitigated during the measurement acquisition and signal-processing stages. ~\cite{trackingsystem,ignore_clutter1,ignore_clutter2,srikar}.

% ~\cite{trackingsystem,ignore_clutter1,ignore_clutter2,srikar}.

%% file: chapter/data_model.tex
\section{Data Model}
\label{sec: data model}
% \subsection{Network Setup}
% % About the network 
% We consider a network consisting of $N$ omnidirectional monostatic radar sensors node with known positions $\mathbf{p}_n = [p_{x_n}, p_{y_n}]^\top \text{, }\forall n = 1, 2,\cdots, N$. The network estimates target position and velocity with noisy range and Doppler measurements gathering from each radar sensor. The sensor network is indicated as $\mathcal{F}$, an undirected graph $\mathcal{F} = (\mathcal{V},\mathcal{N})$, where $\mathcal{V}$ is a set of sensor nodes, $\mathcal{N}$ is a communication link within the network. The neighbors of the $n$ th node are indicated by $j \in \mathcal{N}_n$. 

% % About each node in the network
% Each radar node transmits Linear Frequency Modulated (LFM) signals defined by bandwidth (B), wavelength ($\lambda$) sending $L$ pulses in each burst, where a brust is a series of pulses sent together to enhance signal accuracy, within a Coherent Processing Interval (CPI). The total duration of measuring one estimation of the target is $M\times t_{\text{CPI}}$, with $M$ indicating the number of brust, each providing $M$ range and Doppler shift measurements to localize the target.

\subsection{Signal Model}
We consider each radar node in the network to independently receive noisy measurements, whose statistics depend on environment and the geometry of the radar network~\cite{radar_geometry_placement}. We assume the raw signals are processed and the target is detected ~\cite{radar_principle}, which is a reasonable assumption in high-SNR scenarios~\cite{van2001detection}. Assuming the initial phase and amplitude are unknown, we characterize the noise variance at the radar node $n$, denoted by $\sigma_n^2$ and $\rho_n^2$, as a function of the $n$th node of signal-to-noise ratio ($\text{SNR}_n$), bandwidth $B_n$, number of pulses $L$ and sampling period $t$, i.e.,  
\begin{subequations}\label{eq:signal model}
\begin{align}
\sigma_n^2 
&\ge 
\frac{3c^2}
{8\pi^2 B_n^2 \mathrm{SNR}_n},
\label{eq:sigma model range} \\[0.5em]
\rho_n^2 
&\ge 
\frac{3}
{2\pi^2 t^2 L(L^2-1)\mathrm{SNR}_n}
\approx
\frac{3}
{2\pi^2 t^2 L^3 \mathrm{SNR}_n},
\label{eq:sigma model doppler} \\[0.5em]
\gamma_{n}
&=
\eta \sigma_{n}\rho_{n}\text{,}
\label{eq:sigma model cross term}
\end{align}
\end{subequations}
where the approximation of $\rho_n^2$ holds for large $L$. The coupling between the range and Doppler is modeled as $\gamma_{n} = \eta \sigma_{n} \rho_{n}$, where $\eta$ denotes the correlation coefficient between the range and Doppler shift measurements.
\subsection{Measurement Model}
We denote the target state parameter of an unknown target at time $k$ as $\bsym{\theta}_k = [\bsym{p}_k, \bsym{v}_k]^\top \in \mathbb{R}^{4\times1}$, where~$\bsym{p}_k = [x_k,y_k]^\top$ and $\bsym{v}_k = [\dot x_k, \dot y_k]^\top$ represent the position and velocity of the target in a two-dimensional Cartesian plane. The position of fixed-location $n$th radar sensor node is given by $\bsym{g}_n = [x_n, y_n]^\top$. 
For the $n$th radar node, the noiseless range and Doppler shift measurements at time $k$ are defined as
\begin{subequations}\label{eq: mea model}
\begin{align}
    \label{eq:mea model range}
    % r_{n}(\bsym{\theta}_k) &= \sqrt{(x_k-x_n)^2 + (y_k - y_n)^2}\text{,}\\
        r_{n}(\bsym{\theta}_k) &= ||\bsym{p}_k-\bsym{g}_n||_2\text{,}\\
    % f_{n}(\bsym{\theta}_k) &= \frac{\bsym{v}_k^\top}{\lambda}\cdot \frac{\bsym{p}_n - \bsym{p}_k}{\lvert\lvert\bsym{p}_n - \bsym{p}_k\lvert\lvert}\text{,}
     \label{eq:mea model doppler}
    f_{n}(\bsym{\theta}_k) &= \frac{2\bsym{v}_k^\top}{\lambda}\cdot \frac{    \bsym{p}_k-\bsym{g}_n}{|\bsym{p}_k-\bsym{g}_n|}\text{,}
\end{align}
\end{subequations}
where unit vector $\frac{\bsym{p}_k - \bsym{g}_n}{\lvert\lvert\bsym{p}_k - \bsym{g}_n\lvert\lvert}$ points from the $n$~th radar node to the target and determines the radial component of the target velocity with respect to node $n$.
% % Move this paragraph to MLE / MAP part. Would be easier for readability
% Over $K$ tracking time, each node $n$ collects a sequence of range and Doppler measurements $\{\hat{r}_{n,m}, \hat{f}_{n,m}\}, m = 1\cdots K$, which can be written as the sum of the corresponding noiseless quantities and measurement errors
% \begin{equation}
% \begin{bmatrix}
% \hat{r}_{n,1} \\
% \hat{f}_{n,1} \\
% \vdots \\
% \hat{r}_{n,K} \\
% \hat{f}_{n,K} \\
% \end{bmatrix}
% =
% \begin{bmatrix}
% r_{n}(\bsym{\theta}_1) \\
% f_{n}(\bsym{\theta}_1) \\
% \vdots \\
% r_{n}(\bsym{\theta}_K) \\
% f_{n}(\bsym{\theta}_K)\\
% \end{bmatrix}
% +
% \begin{bmatrix}
% e_{rn,1} \\
% e_{fn,1} \\
% \vdots \\
% e_{rn,K} \\
% e_{fn,K}
% \end{bmatrix}\text{,}
% \end{equation}
% where $\hat{r}_{n,m}$ and $\hat{f}_{n,m}$ are the range and Doppler shift measurements at the $n$ th node and $m$ th measurements, and $e_{rn,m}, e_{fn,m}$ denote the corresponding measured errors. 
For simplicity, we denote all the range-Doppler measurements at time $k$ from the $n$th node as
\begin{equation}
\label{eq:noisy mea model}
    \mbf{z}_{n,k} = \bsym{\mu}_{n}(\bsym{\theta}_k) + \mbf{e}_{n,k}\text{,}
\end{equation}
where $\bsym{\mu}_{n}(\bsym{\theta}_k) = [r_n{(\bsym{\theta}_k)},f_n(\bsym{\theta}_k)]^\top \in \mathbb{R}^{2}$ is the noiseless measurements of the $n$th radar node on target state $\bsym{\theta}_k$ at time $k$, and $ \bsym{e}_{n,k}$ is assumed to be underlying zero-mean Gaussian noise on the measurements of the $n$ th radar node ($\mathbf{e}_{n,k}\sim\mathcal{N}~(0,\mathbf{\Sigma}_{n,k}), \mathbf{\Sigma}_{n,k}\succ 0 $).
The covariance over all $K$ time at node $n$ is assumed to be uncorrelated over time and modeled by
\begin{align}
      \bsym{\Sigma}_n &= \text{blkdiag}(\bsym{\Sigma}_{n,1},\bsym{\Sigma}_{n,2},\cdots,\bsym{\Sigma}_{n,K}) \text{,}
\end{align}
where each 2$\times$2 block $\bsym{\Sigma}_{n,k}$ contains covariance of range and Doppler noise at time $k$. The notation extends the signal-dependent noise model introduced in \eqref{eq:signal model} to a time-indexed representation.
\begin{equation}
\label{eq: mea cov mtx}
        \boldsymbol{\Sigma}_{n,k} = \begin{bmatrix}
        \sigma^2_{n,k} & \gamma_{n,k}\\
        \gamma_{n,k}   & \rho^2_{n,k}       
    \end{bmatrix}  \text{.}
\end{equation}
The entries of $\bsym{\Sigma}_{n,k}$ are related to the signal model given in (\ref{eq:signal model}), and we assume this is  symmetric and positive semi-definite matrix. We further assume that range and Doppler errors in are correlated and jointly follow a bivariate normal distribution with zero mean~\cite{radar_principle}. 
In summary, each sensor node $n$ observes the radial position and velocity~\eqref{eq: mea model}, and  by combining information from all nodes in the network, for a single target in 2 dimensions we aim to estimate the position and velocity vectors $\bsym{\theta}_k = [\bsym{p}_k, \bsym{v}_k]^\top \in \mathbb{R}^{4\times1}$ of a single target in 2D dimension at each time instance $k\leq K$.

%% file: chapter/fusion_algor.tex
\section{Centralized Target Tracking}
\label{sec: centralizede fusion}
This section presents algorithms for distributed radar target tracking, starting from centralized MLE, MAP and EKF baselines.
\subsection{Centralized MLE (C-MLE)}
To estimate the parameter vector $\boldsymbol{\theta}_k$, we can employ the MLE to infer the target state from a batch of measurements~\cite{srikar}.
% Move this paragraph to MLE / MAP part. Would be easier for readability
Specifically, at tracking time step $k$, each node $n$ collects $M$ noisy range and Doppler measurements $[\hat{\mbf{r}}_{n,k},\hat{\mbf{f}}_{n,k}]^\top$ (obtained from (\ref{eq:noisy mea model})) which is a block of $M$ measurement before track time $k$. The node-wise measurements are then transmitted to the fusion center, resulting in the aggregated measurements $\bar{\mbf{z}}_k= [ [\hat{\mbf{r}}_{1,k},\hat{\mbf{f}}_{1,k}]^\top,\cdots,[\hat{\mbf{r}}_{N,k},\hat{\mbf{f}}_{N,k}]^\top]^\top$. The $M$ measurements from $N$ nodes can be written as the sum of the corresponding noiseless quantities and measurement errors
\begin{equation}
\begin{bmatrix}
\hat{r}_{1,k-M+1} \\
\hat{f}_{1,k-M+1} \\
\vdots \\
\hat{r}_{N,k} \\
\hat{f}_{N,k} \\
\end{bmatrix}
=
\begin{bmatrix}
r_{1}(\bsym{\theta}_{k}) \\
f_{1}(\bsym{\theta}_{k}) \\
\vdots \\
r_{N}(\bsym{\theta}_k) \\
f_{N}(\bsym{\theta}_k)\\
\end{bmatrix}
+
\begin{bmatrix}
e_{r1,k-M+1} \\
e_{f1,k-M+1} \\
\vdots \\
e_{rN,k} \\
e_{fN,k}
\end{bmatrix}\text{,}
\end{equation}
where $e_{rn,m}, e_{fn,m}$ denote the corresponding measured errors. Note that the duration of measurement acquisition is much shorter than the tracking interval, i.e. $M\times t_{\text{CPI}} \ll \Delta k$, such that the target motion within each measurement batch is negligible, thereby reducing motion-induced estimation errors. Assuming that aggregated measurements $\bar{\mathbf{z}}_{k}$ are conditionally independent across nodes given $\boldsymbol{\theta}_k$. The C-MLE estimate $\hat{\boldsymbol{\theta}}_k^{\text{C-MLE}}$ is obtained by minimizing the negative log-likelihood function:
\begin{equation}
\label{eq:mle}
    \hat{\boldsymbol{\theta}}_k^{\text{C-MLE}} = \arg \min_{\boldsymbol{\theta}_k}\,\,-\ln p(\bar{\mathbf{z}}_k; \boldsymbol{\theta}_k),
\end{equation}
where the log-likelihood can be factorized as the sum of log likelihood function i.e., $-\ln p(\bar{\mathbf{z}}_k; \boldsymbol{\theta}_k) \sum_{n=1}^{N} l_n(\boldsymbol{\theta}_k)$ consisting of local measurements $\mbf{z}_{n,k}$. The local cost term at each node $n$, ignoring a constant, is defined as
\begin{equation}
\label{eq:local lld}
    l_n(\boldsymbol{\theta}_k) \propto \Bigl(\mathbf{z}_{n,k} - \boldsymbol{\mu}_n(\boldsymbol{\theta}_k)\Bigr)^\top \boldsymbol{\Sigma}_{n,k}^{-1} \Bigl(\mathbf{z}_{n,k} - \boldsymbol{\mu}_n(\boldsymbol{\theta}_k)\Bigr).
\end{equation} 
\subsection{Centralized MAP (C-MAP)}
The maximum a posteriori approach extends MLE by incorporating prior information of the parameters $\bsym{\theta}_k$. When tracking a target over time, the target state $\bsym{\theta}_{k}$ is typically related to the previously predicted state $\hat{\bsym{\theta}}_{k-1}$. Assuming a first order Markov process, the C-MAP estimator is given by 
\begin{equation}
\begin{aligned}
    \hat{\bsym{\theta}}_k^{\text{C-MAP}} 
    &= \argmin_{\bsym{\theta}_k} \,\,-\ln\Bigl(p(\bsym{\theta}_k|\bar{\mbf{z}}_k,\hat{\bsym{\theta}}_{k-1})\Bigr)\text{,}   \\
    &= \argmin_{\bsym{\theta}_k} \,\,-\ln\Bigl(p(\bar{\mbf{z}}_k|\bsym{\theta}_k)\Bigr) - \ln\Bigl(p(\bsym{\theta}_{k}| \hat{\bsym{\theta}}_{k-1})\Bigr)\text{,}  \\
    &= \argmin_{\theta_{k}}\Bigl( \lvert\lvert\bar{\mbf{z}}_{k} - \bar{\bsym{\mu}}(\bsym{\theta}_{k})\rvert\rvert^2_{\bar{\bsym{\Sigma}}_{k}^{-1}} + \\
    &\qquad\qquad\qquad\lvert\lvert\bsym{\theta}_{k} - \hat{\bsym{\theta}}_{n,k|k- 1}\rvert\rvert^2_{\mbf{P}_k^{-1}}\Bigr)\text{,}
\end{aligned}
\end{equation}
 where the likelihood term $p(\bar{\mbf{z}}_k|\bsym{\theta}_k)$ is the joint likelihood as shown in (\ref{eq:local lld}), and $p(\bsym{\theta}_{k}|\bsym{\theta}_{k-1})$ is the prior distribution of the target state from each sensor node  
\begin{equation}p(\bsym{\theta}_{k}|\bsym{\theta}_{k-1}) = \mathcal{N}(\bsym{\theta}_k|\bsym{\hat{\theta}}_{k|k-1},\mathbf{P}_k)\text{,}
 \end{equation}
where $\mathbf{P}_k = \mathbb{E}\left[
\left(\bsym{\theta}_k-\mathbb{E}[\bsym{\theta}_k]\right)
\left(\bsym{\theta}_k-\mathbb{E}[\bsym{\theta}_k]\right)^{\top}
\right]\succ0$ is the state covariance matrix.
% Give update rule of the state cov. 
% In the end give the wls opti problem
 
\subsection{Centralized Extended Kalman filter (C-EKF)}
\label{sec: CEKF}
As an extension of the MLE and MAP estimators, we next consider a recursive Kalman-filter-based approach. Assuming that the target dynamics are known and modeled in state-space form, the Kalman filter recursively updates both the state estimate and its covariance through prediction and correction steps. Compared with MAP estimation, which typically requires solving an optimization problem at each time step, this recursive formulation enables efficient online tracking while explicitly maintaining state uncertainty.

In the centralized Kalman filter setting, a fusion center collects measurements from all radar nodes at each time step $k$ and computes a global state estimate using the known dynamical model $\mbf{f}(\cdot)$ and the range-Doppler measurement model $\bar{\bsym{\mu}}(\cdot)$, where $\bar{\bsym{\mu}}(\cdot) = [\cdots,\mu_{n}(\cdot)^\top,\cdots]^\top$. To this end, we assume the radar network observe a target with state space model
\begin{equation}
\label{eq: CKF-SSM}
\begin{aligned}
    \bsym{\theta}_{k | k-1} &= \mbf{f}(\bsym{\theta}_{k-1 | k-1}) + \mbf{w}_{k}\text{,}\,\mbf{w}_{k}\sim \mathcal{N}(\mbf{0},\mbf{Q}_{k})\text{,}\\
    \bar{\mbf{z}}_{k} &= \bar{\bsym{\mu}}(\bsym{\theta}_{k|k-1}) + \bar{\mbf{e}}_{k}\text{,}\, \bar{\mbf{e}}_{k} \sim \mathcal{N}(\mbf{0},\bar{\bsym{\Sigma}}_{k}) \text{,}
\end{aligned}
\end{equation}
where process noise $\mbf{w}_{k}$ is a zero-mean Gaussian sequence with covariance satisfying $\mbf{Q} = \mathbb{E}[\mbf{w}_{k} \mbf{w}_{k}^\top]\succ 0 $. Likewise, $\bar{\mbf{e}}_{k}$ is a zero-mean Gaussian sequence with covariance satisfying $\bar{\bsym{\Sigma}}_{k} = \mathbb{E}[\bar{\mbf{e}}_{k}\bar{\mbf{e}}_{k}^\top]\succ 0$, where  $\bar{\bsym{\Sigma}}_{k} = \text{blkdiag}(\bsym{\Sigma}_1,\cdots,\bsym{\Sigma}_N)$.

\textbf{Prediction:} For practical implementation for estimates $\bsym{\theta}_{k|k-1}$, we employ the EKF approach, which linearized the dynamics model and measurement model. Given $(\bsym{\theta}_{k-1|k-1}, \mbf{P}_{k-1|k-1})$, the predicted mean
$\bsym{\theta}_{k|k-1} = \mbf{F}\bsym{\theta}_{k-1|k-1}$, and the state covariance is
$\mbf{P}_{k|k-1} = \mbf{F}_k\mbf{P}_{k-1|k-1}\mbf{F}_k^\top + \mbf{Q}_k$, where
$\mbf{F}_k = \left.\frac{\partial \mbf{f}(\bsym{\theta})}{\partial \bsym{\theta}}\right|_{\bsym{\theta}=\bsym{\theta}_{k-1|k-1}}$ is the Jacobian matrix of the dynamics function.

\textbf{Correction:}
In the correction step, measurement model is linearized at $\bsym{\theta}_{k|k-1}$, obtaining
$\bar{\mbf{H}}_k = \left.\frac{\partial \bar{\bsym{\mu}}(\bsym{\theta})}{\partial \bsym{\theta}}\right|_{\bsym{\theta}=\bsym{\theta}_{k|k-1}}$.
The innovation $\bar{\mbf{y}}_k = \bar{\mbf{z}}_k - \bar{\bsym{\mu}}(\bsym{\theta}_{k|k-1})$, and Kalman gain
$\bar{\mbf{S}}_k = \bar{\mbf{H}}_k\bsym{\mbf{P}}_{k|k-1}\bar{\mbf{H}}_k^\top + \bar{\bsym{\Sigma}}_k,$
$\bar{\mbf{K}}_k = \mbf{P}_{k|k-1}\bar{\mbf{H}}_k^\top \bar{\mbf{S}}_k^{-1}$ respectively.
Finally, the posterior mean and covariance are updated as
\begin{equation}
\begin{aligned}
\label{eq: CKF-correction}
\hat{\bsym{\theta}}_{k|k}^\text{C-EKF} &= \hat{\bsym{\theta}}_{k|k-1} + \bar{\mbf{K}}_k\bar{\mbf{y}}_k,\\
\mbf{P}_{k|k}^{\text{C-EKF}} &= \mbf{P}_{k|k-1} - \bar{\mbf{K}}_k\bar{\mbf{H}}_k\mbf{P}_{k|k-1} .
\end{aligned}
\end{equation}
  
 In the centralized sensor fusion setting, all local measurements $\mbf{z}_{n,k}$, $\bsym{\mu}_n({\bsym{\theta}_k})$ and $\bsym{\Sigma}_{n,k}$ are transmitted to a fusion center, which computes C-MLE, C-MAP or C-EKF to determine the target position and velocity from range and Doppler measurements. However, these centralized frameworks suffer from single point failure and do not scale with the network size due to bandwidth limit. To overcome these limitations, we now propose a distributed framework that distributes the computation across sensor nodes.
\section{Distributed Target Tracking}
\label{sec: distributed fusion}
This section presents distributed tracking algorithms for distributed MLE, MAP and EKF.
\subsection{Distributed MLE~(D-MLE)}
We consider a monostatic radar sensor network as a connected undirected graph. The neighbors of $n$th node are denoted by $\mathcal{N}_n$, where $|\mathcal{N}_n| \geq 2$ since at least two neighbor radar nodes are needed to jointly-localize a target in two-dimensional space with only radial range measurements~\cite{triangluar_properties}. In our prior work~\cite{srikar}, a distributed maximum likelihood estimation (D-MLE) method was proposed for target localization. This approach can be extended to target tracking by aggregating $M$ measurements before each tracking time step $k$. The local negative log-likelihood is then minimized using measurements from each node and its neighbors, while consensus among neighboring estimates is enforced through edge variables
\begin{equation}
\label{eq: MLE ADMM}
\begin{aligned}
&\hat{\bsym{\theta}}_{n,k}^{\text{D-MLE}} = \argmin_{\bsym{\theta}_{n,k}}\Bigl( \lvert\lvert\mbf{z}_{n,k} - \bsym{\mu}_n(\bsym{\theta}_{n,k})\rvert\rvert^2_{\bsym{\Sigma}_{n,k}^{-1}} \\
&\qquad\qquad\qquad+\sum_{j \in \mathcal{N}_n}\lvert\lvert\mbf{z}_{j,k} - \bsym{\mu}_j(\bsym{\theta}_{n,k})\rvert\rvert^2_{\bsym{\Sigma}_{j,k}^{-1}} \Bigl)\text{,}\\
    &\textrm{s.t.} \quad \bsym{\theta}_{n,k} = \bsym{\vartheta}_{n,j}\text{,} \, \bsym{\theta}_{j,k} = \bsym{\vartheta}_{n,j} \, \forall j \in \mathcal{N}_n
\text{,}
\end{aligned}
\end{equation}
where $\bsym{\vartheta}_{n,j} \in \mathbb{R}^{4\times1}$ is a auxiliary constraint variable for node~$j \in \mathcal{N}_n$. 
\subsection{Distributed MAP~(D-MAP)}
Building on the D-MLE approach, for the $n$th radar node, the local MAP estimator objective function is sum of the likelihood~\eqref{eq:local lld} plus the prior term and the posterior terms associated with its neighboring nodes, given by
\begin{equation}
\begin{aligned}
\label{eq: MAP dectrl.}
\ln{(p_n (\bsym{\theta}_{n,k}|\mbf{Z}_{n,k}) }) = l_n(\bsym{\theta}_{n,k})+\ln p(\bsym{\theta}_{n,k} | \bsym{\theta}_{n,k-1}) + \\ \sum_{j\in\mathcal{N}_n}l_j(\bsym{\theta}_{n,k}) + \ln p(\bsym{\theta}_{n,k} | \bsym{\theta}_{j,k-1})\text{,}    
\end{aligned}
\end{equation}
where all the collected measurements at the $k$th instant at the $n$th node are denoted by $\mathbf{Z}_{n,k}$ = $\{\mathbf{z}_{n,k}\} \cup \{\mathbf{z}_{j,k}| j \in \mathcal{N}_n\}$ , including its own measurements and those received from its neighbors. In contrast to the centralized solution, which relies on a single fusion center, we introduce edge variables to enforce the agreement enables each pair of connected nodes $(n,j)$ to enforce local consensus constraints, ensuring their estimates converge to a common value without the need for high-bandwidth data transmission to a central node. The optimization problem at time $k$ is then formulated as minimizing the negative local log posterior at each node, subject to equality constraints between the node variables and the corresponding edge variables 
\begin{equation}
\label{eq: MAP ADMM}
\begin{aligned}
&\hat{\bsym{\theta}}_{n,k}^{\text{D-MAP}} = \argmin_{\bsym{\theta}_{n,k}}\Bigl( \lvert\lvert\mbf{z}_{n,k} - \bsym{\mu}_n(\bsym{\theta}_{n,k})\rvert\rvert^2_{\bsym{\Sigma}_{n,k}^{-1}} +\\ &\qquad\qquad\qquad\qquad\,\,\lvert\lvert\bsym{\theta}_{n,k} - \bsym{\theta}_{n,k-1}\rvert\rvert^2_{\mbf{P}^{-1}_{n,k}}\\
&+\sum_{j \in \mathcal{N}_n}( \lvert\lvert\mbf{z}_{j,k} - \bsym{\mu}_j(\bsym{\theta}_{n,k})\rvert\rvert^2_{\bsym{\Sigma}_{j,k}^{-1}} + \lvert\lvert\bsym{\theta}_{n,k} - \bsym{\theta}_{j,k-1}\rvert\rvert^2_{\mbf{P}_{j,k}^{-1}})\Bigr)\text{,}\\
    &\textrm{s.t.} \quad \bsym{\theta}_{n,k} = \bsym{\vartheta}_{n,j}\text{,} \quad \bsym{\theta}_{j,k} = \bsym{\vartheta}_{n,j} \quad \forall j \in \mathcal{N}_n
\text{,}
\end{aligned}
\end{equation}
where $\bsym{\vartheta}_{n,j} \in \mathbb{R}^{4\times1}$ is an auxiliary constraint variable for node~$j \in \mathcal{N}_n$.
\begin{figure*}[b]
\hrule
\begin{eqnarray}
\label{eq: primal theta}
    \bsym{\theta}_{n,k}(i+1) &=&\argmin_{\bsym{\theta}_{n,k}}\left(-\ln{p_n(\bsym{\theta}_{n,k};\mbf{Z}_{n,k})}+
    \sum_{j \in \mathcal{N}_n} \Bigl( \bsym{\psi}_{nj}^T(i)(\bsym{\theta}_{n,k} - \bsym{\vartheta}_{nj}(i))+ \left\lVert\frac{1}{2}(\bsym{\theta}_{n,k} - \bsym{\vartheta}_{nj}(i)) \right\rVert_{\bsym{\Phi}}^2 \Bigr)\right)\text{,} \\
\label{eq: primal consensus var}
    \bsym{\vartheta}_{nj}(i+1) 
    &=& \frac{1}{2}\Bigl[\bsym{\Phi}^{-1}\Bigl(\bsym{\psi}_{nj}(i)+\bsym{\psi}_{jn}(i)\Bigr)+\bsym{\theta}_{n,k}(i+1)+\bsym{\theta}_{j,k}(i+1)\Bigr]\text{,}\quad j \in \mathcal{N}_n\text{,} \\
\label{eq: dual update}
    \bsym{\psi}_{nj}(i+1) &=& \bsym{\psi}_{nj} (i) + \bsym{\Phi}\Bigl(\bsym{\theta}_n(i+1)-\bsym{\bsym{\vartheta}}_{nj}(i+1)\Bigr)\text{,}\quad j \in \mathcal{N}_n \text{,}
\end{eqnarray}
\end{figure*}

The optimization problem can then be solved using a consensus ADMM framework~\cite{consensus_ADMM_on_network}. To accurately evaluate the behavior of the formulation, this study relies only on the standard consensus ADMM. While we acknowledge that variants like proximal ADMM~\cite{PxADMM} and PDMM~\cite{pdmm} offer specific computational benefits, our priority is to thoroughly understand the baseline convergence properties of the MAP formulation before exploring those extensions.
%% Stop criteria
\begin{equation}
\label{eq: primal residual}
    \delta_n(i+1) = \sqrt{\sum_{j\in \mathcal{N}_n}\vert\vert\bsym{\theta}_{n,k}(i+1) - \bsym{\vartheta}_{n,j}(i+1)\vert\vert^2_2}\text{,}
\end{equation}
\begin{equation}
\label{eq: dual residual}
    \epsilon_n(i+1) = \sqrt{\sum_{j\in \mathcal{N}_n}\vert\vert\bsym{\psi}_{nj}(i+1)  - \bsym{\psi}_{nj}(i) \vert\vert^2_2}\text{.}
\end{equation}
The algorithm solves consensus optimization at each time step $k$. At each optimization iteration $i$, the update equations are given in (\ref{eq: primal theta}), (\ref{eq: primal consensus var}) and (\ref{eq: dual update}), 
where dual variable $\bsym{\psi}\in \mathbb{R}^{4\times1}$ and $\bsym{\Phi}\in\mathbb{R}^{4\times4}$ and $\bsym{\Phi}\succ 0$ is a predefined penalty matrix which is invertible and positive semi-definite. Using different penalties for position and velocity is necessary for balancing their relative influence in the optimization and achieving stable convergence of the consensus updates. At each optimization iteration $i$, we evaluate convergence by calculating primal residual $\delta_n(i)$ and dual residual $s_n(i)$

The primal residual shows the discrepancies between local estimates and neighboring ones and the dual residual measures the change in the dual variable between iterations.The proposed D-MAP for target tracking is summarized in Algorithm~\ref{alg:MAP_ADMM}.
%% Consensus ADMM algorithm
\input{algorithm/MAP_consensusADMM}
\subsection{Distributed EKF (D-EKF)}
% Why need DKF?
In addition to the batch-processing distributed estimators (D-MLE and D-MAP), we also consider a recursive approach for target tracking in a distributed setting. The batch estimators do not explicitly exploit a target-dynamics model, so the state covariance $\mbf{P}_k$ is not updated for every time step in~(\ref{eq: MAP ADMM}). In contrast, a recursive estimator assumes that the target dynamics is known and uses them to propagate the state over time. Incorporating this dynamical prior improves estimation accuracy and enables faster convergence, particularly when measurements are noisy. \\
\textbf{State Space Model:} Similar to the centralized state-space model in (\ref{eq: CKF-SSM}), each radar node maintains a local state-space model to estimate the target state from its range–Doppler measurements while exploiting the assumed target dynamics. In the distributed setting, however, the prediction and measurement models are evaluated locally at each radar node
\begin{equation}
\label{eq: SSM}
\begin{aligned}
    \bsym{\theta}_{n,k | k-1} &= \mbf{f}(\bsym{\theta}_{n,k-1 | k-1}) + \mbf{w}_{n,k}\text{,}\,\mbf{w}_{n,k}\sim \mathcal{N}(\mbf{0},\mbf{Q}_{n,k})\text{,}\\
    \mbf{z}_{n,k|k} &= \bsym{\mu}_{n}(\bsym{\theta}_{n,k|k-1}) + \mbf{e}_{n,k}\text{,}\, \mbf{e}_{n,k} \sim \mathcal{N}(\mbf{0},\bsym{\Sigma}_{n,k}) \text{,}
\end{aligned}
\end{equation}
where $\bsym{\theta}_{n,k | k-1}$ denotes the node-dependent predicted state at node $n$ and time $k$ given the locally estimated posterior state at the previous time step $k-1$. 
%About F and M
Similar to C-EKF, the dynamical function $\mbf{f}(\cdot)$ is assumed to be globally known and shared across all radar nodes. However, the global estimates are  not available, thus the Jacobian $\mbf{F}_{n,k} = \left.\frac{\partial \mbf{f}}{\partial \bsym{\theta}}\right|_{\bsym{\theta}=\bsym{\theta}_{n,k-1|k-1}}$ is linearized with the local estimates $\bsym{\theta}_{n, k-1|k-1}$. On the other hand, the measurement model $\bsym{\mu}_n(\cdot)$ is also node-specific which is linearized as $\mbf{H}_{n,k} = \left.\frac{\partial \bsym{\mu}_n}{\partial \bsym{\theta}}\right|_{\bsym{\theta}=\bsym{\theta}_{n,k|k-1}}$.

For prediction step, given the posterior estimate in previous time step $(\bsym{{\theta}}_{n,k-1|k-1}, \mbf{P}_{n,k-1|k-1})$, the EKF propagate the state mean through the dynamics model, compute the predicted mean and covariance 
\begin{equation}
    \label{eq: EKF prediction}
    \begin{aligned}
    \bsym{\theta}_{n,k|k-1} &= \mbf{f}(\bsym{\theta}_{n,k-1|k-1})\text{,}\\
    \mbf{P}_{n,k|k-1} &= \mbf{F}_{n,k}\mbf{P}_{n,k-1|k-1}\mbf{F}_{n,k}^\top +\mbf{Q}_{n,k}\text{.}
    \end{aligned}
\end{equation} \\ 
\textbf{Consensus Optimization}: The correction step of the distributed EKF is different from the standard Kalman filter in (\ref{eq: CKF-correction}). The correction among radar nodes is reformulated as distributed optimization problem with MAP  as objective function\cite{RYU}. The only difference between the distributed MAP and EKF lies on the prior term, the state covariance $\mbf{P}_{n,k|k-1}$ and the predicted estimated $\bsym{\theta}_{n,k|k-1}$ is updated recursively rather than fixed.
\begin{equation}
\label{eq: EKF ADMM}
\begin{aligned}
\hat{\bsym{\theta}}_{n,k|k}^{\text{D-EKF}}
&= \argmin_{\bsym{\theta}_{n,k}}
l_{n,k}
\bigl(\bsym{\theta}_{n,k}),\\
l_{n,k}=&\left\|\mbf{z}_{n,k}-\bsym{\mu}_{n}(\bsym{\theta}_{n,k})\right\|_{\bsym{\Sigma}_{n,k}^{-1}}^{2}
+\left\|\bsym{\theta}_{n,k}-\bsym{\theta}_{n,k|k-1}\right\|_{\mbf{P}_{n,k|k-1}^{-1}}^{2}\\
&+\sum_{j\in\mathcal{N}_n}\Biggl[
\left\|\mbf{z}_{j,k}-\bsym{\mu}_{j}(\bsym{\theta}_{n,k})\right\|_{\bsym{\Sigma}_{j,k}^{-1}}^{2}\\
&\quad\qquad+\left\|\bsym{\theta}_{n,k}-\bsym{\theta}_{j,k|k-1}\right\|_{\mbf{P}_{j,k|k-1}^{-1}}^{2}\Biggr],\\
\text{s.t.}\,&\bsym{\theta}_{n,k}=\bsym{\vartheta}_{n,j},\quad\bsym{\theta}_{j,k}=\bsym{\vartheta}_{n,j},\quad\forall j\in\mathcal{N}_n .
\end{aligned}
\end{equation}
% \begin{equation}
% \label{eq: EKF ADMM}
% \begin{aligned}
% \hat{\bsym{\theta}}_{n,k} &= \argmin_{\bsym{\theta}_{n,k}}\Bigl(- \lvert\lvert\mbf{z}_{n,k} - \bsym{\mu}(\bsym{\theta}_{n,k})\rvert\rvert^2_{\bsym{\Sigma}_{n,k}} -\lvert\lvert\bsym{\theta}_{n,k} - \bsym{\theta}_{n,k|k-1}\rvert\rvert^2_{\mbf{P}_{n,k|k-1}}\\
% &+\sum_{j \in \mathcal{N}_n }(- \lvert\lvert\mbf{z}_{j,k} - \bsym{\mu}(\bsym{\theta}_{j,k})\rvert\rvert^2_{\bsym{\Sigma}_{j,k}} -\lvert\lvert\bsym{\theta}_{j,k} - \bsym{\theta}_{j,k|k-1}\rvert\rvert^2_{\mbf{P}_{j,k|k-1}})\Bigr)\text{,}\\
% &\textrm{s.t.} \quad \bsym{\theta}_{n,k} = \bsym{\vartheta}_{n,j}\text{,} \quad \bsym{\theta}_{j,k} = \bsym{\vartheta}_{n,j}\text{,} \quad \forall j \in \mathcal{N}_n
% \text{,}
% \end{aligned}
% \end{equation}
Similarly, the consensus optimization problem can be solved by (\ref{eq: primal theta}), (\ref{eq: primal consensus var}), and  (\ref{eq: dual update}).

For the covariance update, we follow the similar consensus optimization approach. We define the information matrix as inverse of the state covariance matrix $\bsym{\Omega}_{n,k} \coloneq \mbf{H}_{n,k}^\top \bsym{\Sigma}_{n,k}^{-1}\mbf{H}_{n,k}$, represent the local information contribution from the measurements of the $n$th node. Therefore we formulate a consensus optimization problem 
\begin{equation}
\label{eq:neighbor_consensus_opt}
\begin{aligned}
\hat{\bsym{\xi}}_{n,k}&= \argmin_{\xi_{n,k}}
\sum_{i=\mathcal{N}_n \cup n}
\left\|  \bsym{\xi}_{n,k}-|\mathcal{N}_n+1|\bsym{\omega}_{i,k} 
\right\|^2
\\
&\text{s.t.} \quad
\bsym{\xi}_{n,k} = \bsym{\xi}_{j,k}, 
\quad \forall j\in\mathcal{N}_n,
\end{aligned}
\end{equation}
where $\bsym{\omega}_i = \operatorname{vec}_h(\mbf{H}_{n,k}^\top \bsym{\Sigma}_{n,k}^{-1}\mbf{H}_{n,k}) \in \mathbb{R}^{N_{\text{cov}}}$ and dimension of the vector by half-vectorizing a covariance matrix $N_{\text{cov}} = d(d+1)/2$, $d$ is dimension of the estimate state. Then $\hat{\mbf{\Omega}}_{n,k|k}^{\text{D-EKF}} = \operatorname{vec}_h^{-1}(\hat{\xi}_{n,k})$. This optimization problem can be iteratively solved by dual ascent. Alternatively, to avoid the computational complexity of solving the above optimization problem, we can resort to a suboptimal but computationally efficient approach~\cite{RYU}. In this approach, each node updates its information matrix following 
\begin{equation}
\label{eq: information udpate}
\hat{\bsym{\Omega}}_{n,k|k}^{\text{D-EKF}} = \mathbf{P}_{n,k|k-1}^{-1} +  \sum_{j \in \mathcal{N}_n\cup n}\bsym{\Omega}_{j,k} \text{,}
\end{equation}
where $\bsym{\Omega}_{n,k} = \mbf{H}_{n,k}^\top \bsym{\Sigma}_{n,k}^{-1}\mbf{H}_{n,k}$.
The update covariance is then computed as $\hat{\mbf{P}}_{n,k|k}^{\text{D-EKF}} = (\hat{\bsym{\Omega}}_{n,k}^{\text{D-EKF}})^{-1}$. Although this approach could be a suboptimal solution as compared to the centralized approach, it significantly reduces the computational load, making it suitable for a resource-constrained radar node. The proposed D-EKF for target tracking is summarized in Algorithm~\ref{alg:EKF_ADMM}.

\input{algorithm/EKF_consensusADMM}

% \subsection{Posterior Cram\'er Rao Lower Bound for distributed EKF}
% The parameter of the interest at time $k$  $\boldsymbol{\theta}_k$ is stochastic and with prior distribution $\boldsymbol{\theta}_k \sim p(\boldsymbol{\theta)_k}$. Measurements data $\mathbf{z}$ are sampled from $\boldsymbol{\theta}$, and the Bayes MSE can be expressed as:
% \begin{equation*}
% \text{BMSE}(k) = \mathbb{E}_{\boldsymbol{\theta}}(\hat{\boldsymbol{\theta}}_k-\boldsymbol{\theta}_k)^2
% \end{equation*}

% Assume regulation condition of the BMSE function meet, there exist a lower bound of the BMSE such that (in vector form):

% \begin{align*}
% \mathbb{E}\!\left[(\hat{\boldsymbol{\theta}}-\boldsymbol{\theta})
% (\hat{\boldsymbol{\theta}}-\boldsymbol{\theta})^{\mathsf T}\right]
% \ \succeq\ \mathbf{J}_B^{-1},
% \end{align*}

%% file: algorithm/MAP_consensusADMM.tex
\begin{algorithm}[t]
  \caption{D-MAP for $n$th radar node at $k$th instant}
  \label{alg:MAP_ADMM}
  \begin{algorithmic}[1]
    % \noindent\textbf{Input:} 
    \Require Initial guess of estimates $\bsym{\theta}_{n,k}(0)=\hat{\bsym{\theta}}_{n,k-1}$, tolerance $\varepsilon^{\mathrm{pri}}$, $\varepsilon^{\mathrm{dual}}$, penalty matrix $\mathbf{\Phi}$, neighbor set $\mathcal{N}_n$
    \Ensure $\hat{\bsym{\theta}}_{n,k}^{\text{D-MAP}}$
    \State Initialize $\bsym{\psi}_{n,j}(0)>0$, and $\bsym{\vartheta}_{n,j}(0) > 0\, ,\forall\, j \in \mathcal{N}_n$
    % Define $\boldsymbol{\Phi}$ based on SNR 
    \State Initialize optimization iteration~$i = 0$
    \While{$\delta_n(i+1) > \varepsilon^{\mathrm{pri}}$ or $s_n(i+1) > \varepsilon^{\mathrm{dual}}$}
      \State Solve for $\bsym{\theta}_{n,k}(i+1)$ using (\ref{eq: primal theta}) and prior $\hat{\bsym{\theta}}_{n,k-1}$
      \State Transmit $\bsym{\psi}_{nj}(i)$ to all neighbors $j \in \mathcal{N}_n$
      \State Transmit $\bsym{\theta}_{n,k}(i+1)$ to neighbors $j \in \mathcal{N}_n$
      \State Update $\bsym{\vartheta}_{nj}(i+1)$ using (\ref{eq: primal consensus var})
      \State Transmit $\bsym{\vartheta}_{nj}(i+1)$ to neighbors $j \in \mathcal{N}_n$
      \State Update $\bsym{\psi}_{nj}(i+1)$ using (\ref{eq: dual update})
      \State Calculate $\delta_n(i+1)$, $\epsilon_n(i+1)$ using \eqref{eq: primal residual} and \eqref{eq: dual residual}
      \State Increment i
      % \State Increment $i$
    \EndWhile
    \vspace{+5pt}
    \State $\hat{\bsym{\theta}}_{n,k}^{\text{D-MAP}} = \bsym{\theta}_{n,k}(i)$
  \end{algorithmic}
\end{algorithm}

%% file: algorithm/EKF_consensusADMM.tex
\begin{algorithm}[t]
  \caption{D-EKF for $n$th radar node at $k$th instant}
  \label{alg:EKF_ADMM}
  \begin{algorithmic}[1]
    \Require  Estimate $\hat{\bsym{\theta}}_{n,k-1|k-1}$, state covariance $\hat{\mbf{P}}_{n,k-1|k-1}$, tolerance $\varepsilon^{\mathrm{pri}}$, $\varepsilon^{\mathrm{dual}}$, penalty matrix $\mathbf{\Phi}$, neighbor set $\mathcal{N}_n$
    \Ensure $\hat{\bsym{\theta}}_{n,k|k}^{\text{D-EKF}}$, $\hat{\mbf{P}}_{n,k|k}^{\text{D-EKF}}$
    %% First do the prediction
    \State $\mbf{F}_{n,k} = \left.\frac{\partial \mbf{f}}{\partial \bsym{\theta}}\right|_{\bsym{\theta}=\bsym{\theta}_{n,k-1|k-1}}$ \Comment{Local prediction} 
    % \Comment{Linearized dynamics model}
    \State Update $\bsym{\theta}_{n,k|k-1}$ and $\mbf{P}_{n,k|k-1}$ using \eqref{eq: EKF prediction}
    \State Initialize $\bsym{\psi}_{n,j}(0)>0$, and $\bsym{\vartheta}_{n,j}(0) > 0\, ,\forall\, j \in \mathcal{N}_n$
    \State Initialize optimization iteration~$i = 0$
    \State Initialize $\bsym{\theta}_{n,k}(0) = \bsym{\theta}_{n,k|k-1}$
    \While{$\delta_n(i+1) > \varepsilon^{\mathrm{pri}}$ or $\epsilon_n(i+1) > \varepsilon^{\mathrm{dual}}$}
      \State Update $\bsym{\theta}_{n,k}(i+1)$ using \eqref{eq: EKF ADMM}
      \State Transmit $\bsym{\psi}_{nj}(i)$ to all neighbors $j \in \mathcal{N}_n$
      \State Transmit $\bsym{\theta}_{n,k}(i+1)$ to neighbors $j \in \mathcal{N}_n$
      \State Update $\bsym{\vartheta}_{nj}(i+1)$ using (\ref{eq: primal consensus var})
      \State Transmit $\bsym{\vartheta}_{nj}(i+1)$ to neighbors $j \in \mathcal{N}_n$
      \State Update $\{\bsym{\psi}_{nj}(i+1)\}$ using (\ref{eq: dual update})
      \State Calculate $\delta_n(i+1)$, $\epsilon_n(i+~1)$ using \eqref{eq: primal residual} and \eqref{eq: dual residual}
      \State Increment $i$
    \EndWhile \Comment{Distributed Correction via ADMM}
    \vspace{+5pt}
    \State $\hat{\bsym{\theta}}_{n,k|k}^{\text{D-EKF}} = \bsym{\theta}_{n,k}(i)$
    
    \State $\mbf{H}_{n,k} = \left.\frac{\partial \bsym{\mu}_n}{\partial \bsym{\theta}}\right|_{\bsym{\theta}=\bsym{\theta}_{n,k|k-1}}$ \Comment{Covariance Update}
    % \Comment{Linearized measurement model}
    % \State $\bsym{\Omega}_{n,k} = \mbf{H}_{n,k}^\top \bsym{\Sigma}_{n,k}^{-1}\mbf{H}_{n,k} + \bsym{\Omega}_{n,k|k-1}$
    \State Update $\hat{\bsym{\Omega}}_{n,k|k}$ using \eqref{eq: information udpate}
    \State $\hat{\mbf{P}}_{n,k|k}^{\text{D-EKF}} = \hat{\bsym{\Omega}}_{n,k|k}^{-1}$
  \end{algorithmic}
\end{algorithm}

%% file: chapter/simulation_polish.tex
\section{Simulations}
\label{sec: simulation}
\subsection{Experimental Setup}
\begin{figure}[t]
    \centering
    \includegraphics[width=\linewidth]{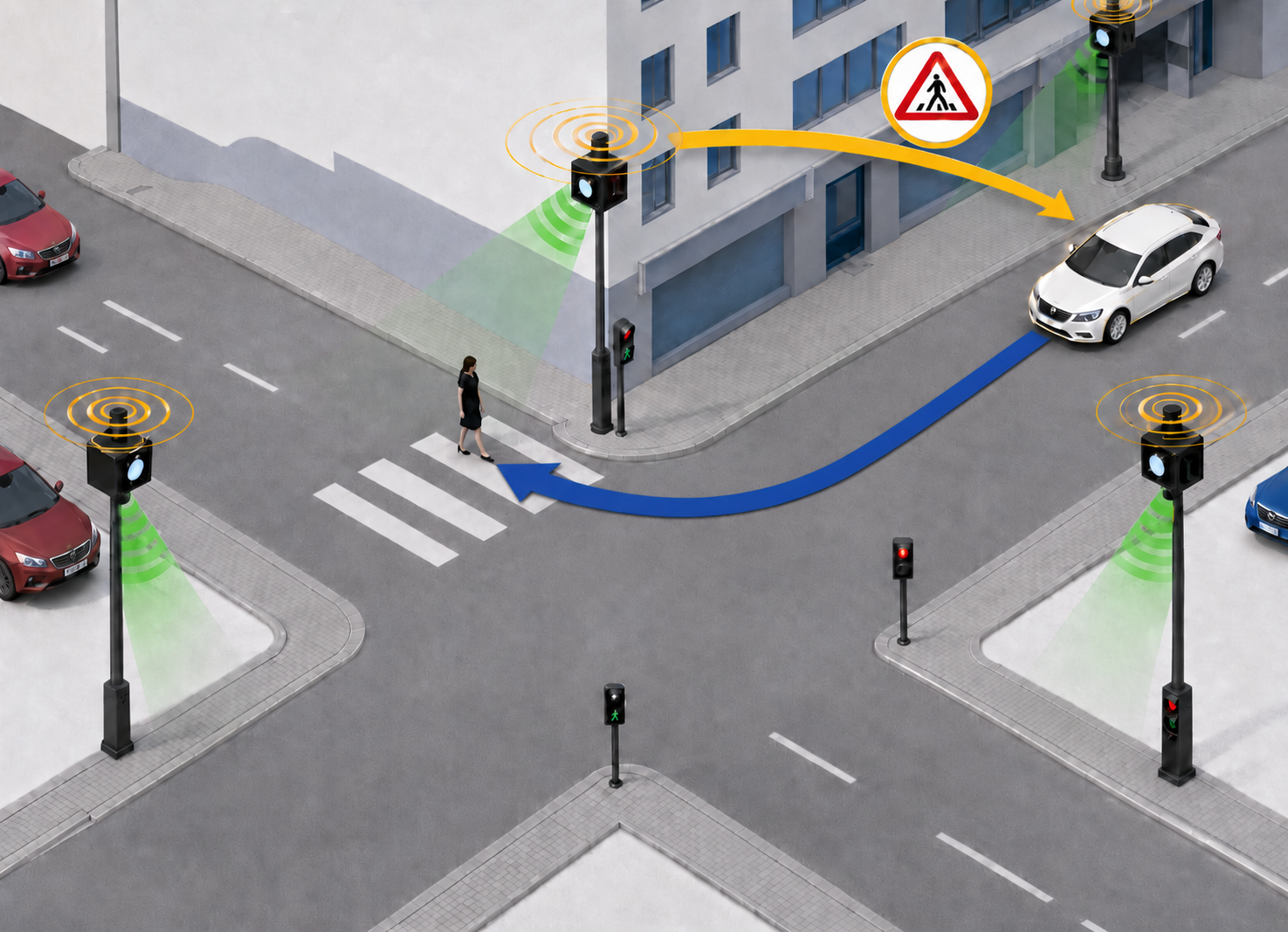}
    \caption{Illustration of a cooperative roadside radar network for target tracking in an urban intersection. Multiple radar nodes (atop a black colored post along the street) observe the moving target (e.g., the pedestrian) from different viewpoints. The illustration is generated using Google Gemini.}
    \label{fig: situation awareness}
\end{figure}

Simulations are conducted to demonstrate and evaluate the proposed distributed radar target-tracking approaches. The considered setup is motivated by cooperative roadside radar sensing in urban intersections, as illustrated in Fig.~\ref{fig: situation awareness}, where multiple radar nodes are arranged in designed geometry, observing a moving target from different viewpoints and exchange information locally to improve tracking robustness. To obtain a controlled and symmetric evaluation setting, we consider ($N=10$) radar nodes deployed uniformly on a circle with a radius of $20$ m. The target starts from ($[-30,-30]^\top$) in the predefined global coordinate frame and moves with a nominal constant speed along an approximately (45$^\circ$) direction. To emulate a more realistic non-straight trajectory, small random perturbations are added to the heading angle over time. Each radar transmits an LFM signal with bandwidth ($B=10\text{MHz}$), number of pulses ($L=64$), and wavelength ($\lambda=0.03$m). We assume time-invariant and node-independent process and measurement noise covariances, i.e.,
$\mbf{\Sigma}_{n,k} = \mbf{\Sigma}$ and $\mbf{Q}_{n,k} = \mbf{Q}$,
for all $n$ and $k$. We use the MATLAB solver $\texttt{fmincon}$ to solve the consensus optimization problem.

For the batch-processing algorithms (MAP and MLE), measurements are collected over $L=64$ pulses before estimating the target state. Consequently, the state estimate is updated once per measurement batch. On the other hand, for the recursive processing approach (EKF), measurements are processed sequentially, and the estimate of ($\bsym{\theta}$) is updated at each time step as new measurements become available. In the simulations, a constant-velocity (CV) dynamics model is employed.

We compare three distributed target tracking algorithm (D-MLE, D-MAP, D-EKF) with centralized algorithm (C-MLE, C-MAP, C-EKF). Each simulation is run over $20$ Monte Carlo runs.  For evaluation, we use the root mean square error (RMSE) of position and velocity
\begin{equation}
\label{eq:rmse_general_component}
\mathrm{RMSE}_{
\phi
}
=
\sqrt{
\frac{1}{N_{\text{MC}}\cdot N \cdot K}
\sum_{q = 1}^{N_{\text{MC}}}\sum_{n=1}^{N}\sum_{k = 1}^{K}
\|
\hat{\bsym{\phi}}_{n,k,q}
-
\bsym{\phi}_{n,k}
\|^2
},
\end{equation} where $\bsym{\phi}$ is the 2D position and velocity, i.e.,
$\bsym{\phi} \in \{\mathbf{p},\mathbf{v}\}$.
The index $q=1,\ldots,N_{\mathrm{MC}}$ denotes the Monte Carlo trial,
$k=1,\ldots,K$ denotes the discrete time step, and $n=1,\ldots,N$ denotes the
radar node. This notation is introduced to accommodate diverse simulation used in this section. In these simulations, we use $N_{\text{MC}} = 20$, $N = 10$, $K = 384$. For D-MLE and D-MAP, the RMSE is averaged over the tracking time steps $k$, Monte Carlo trials $N_{\text{MC}}$, and radar nodes $N$, whereas for D-EKF, the RMSE is evaluated at each time step and averaged only over Monte Carlo trials $N_{\text{MC}}$ and radar nodes $N$.
% The RMSE is summarized as
% $\boldsymbol{\mathrm{RMSE}}=\left[\mathrm{RMSE}_{p},\mathrm{RMSE}_{v}\right]^\top, $ corresponding to the position and velocity components of the target state.

\subsection{Centralized and Distributed Target Tracking using MAP}
\begin{figure}[t]
    \centering
    \includegraphics[width=\linewidth]{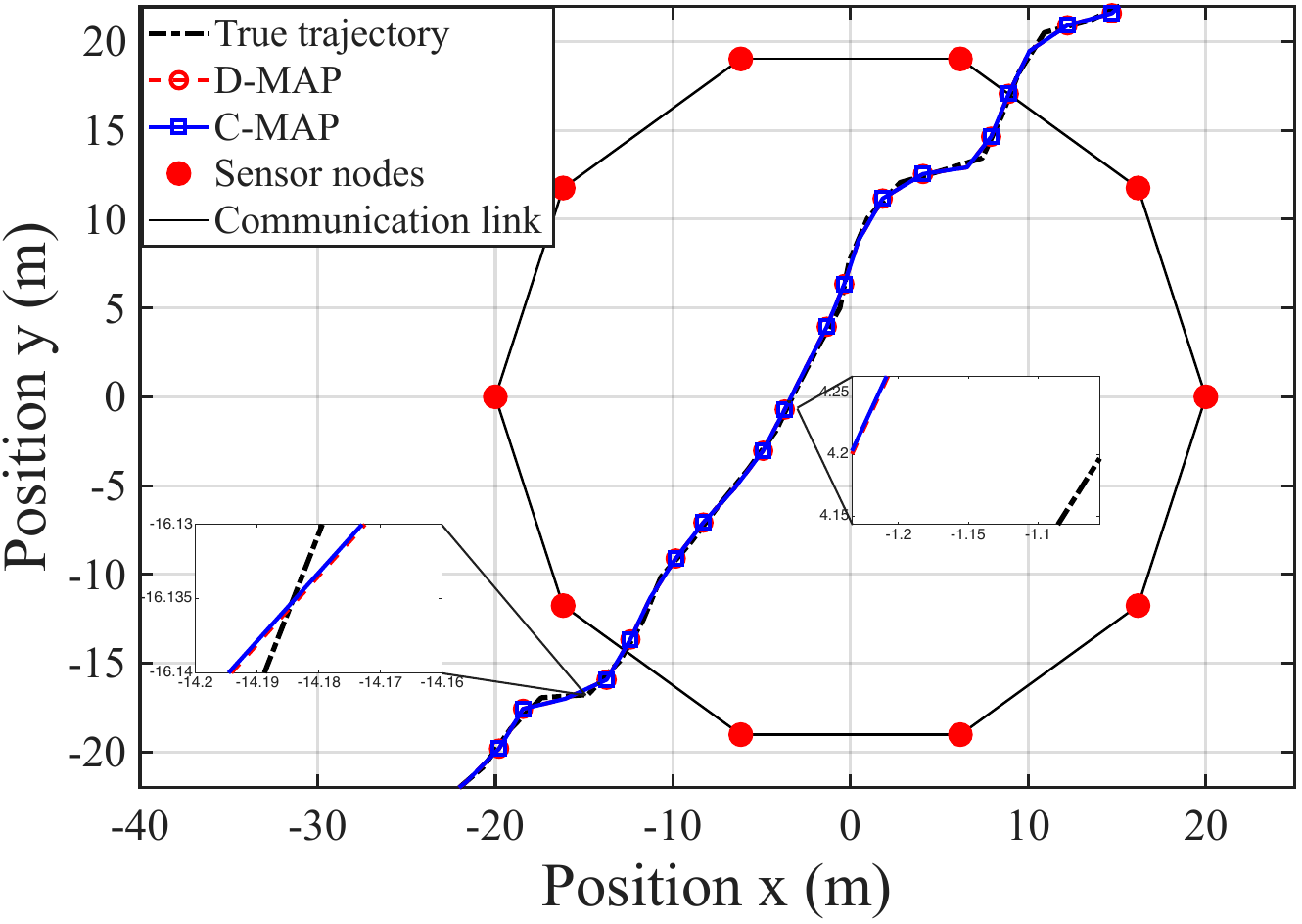}
    \caption{Tracking trajectory comparison for MAP approaches. The inset plot shows the overlap between D-MAP, C-MAP.}
    \label{fig: MAP trajectory}
\end{figure}

Fig. \ref{fig: MAP trajectory} shows the tracking trajectory obtained by MAP approach. Both MAP-based estimators closely follow the true trajectory throughout the tracking interval, demonstrating that the range-Doppler measurements collected from the radar network provide sufficient information. The close overlap between D-MAP and C-MAP indicates that the proposed distributed MAP formulation achieves near-centralized performance using only neighbor-to-neighbor communication. The zoomed inset highlights the distributed solution remains close to the centralized estimate locally, with only a small deviation from the ground-truth trajectory. Although the distributed approach may incur a slight loss in accuracy compared to the centralized solution, it offers improved robustness and scalability by eliminating reliance on a fusion center.
\subsection{Centralized and Distributed Target Tracking using EKF}
\begin{figure}[t]
    \centering
    \includegraphics[width=\linewidth]{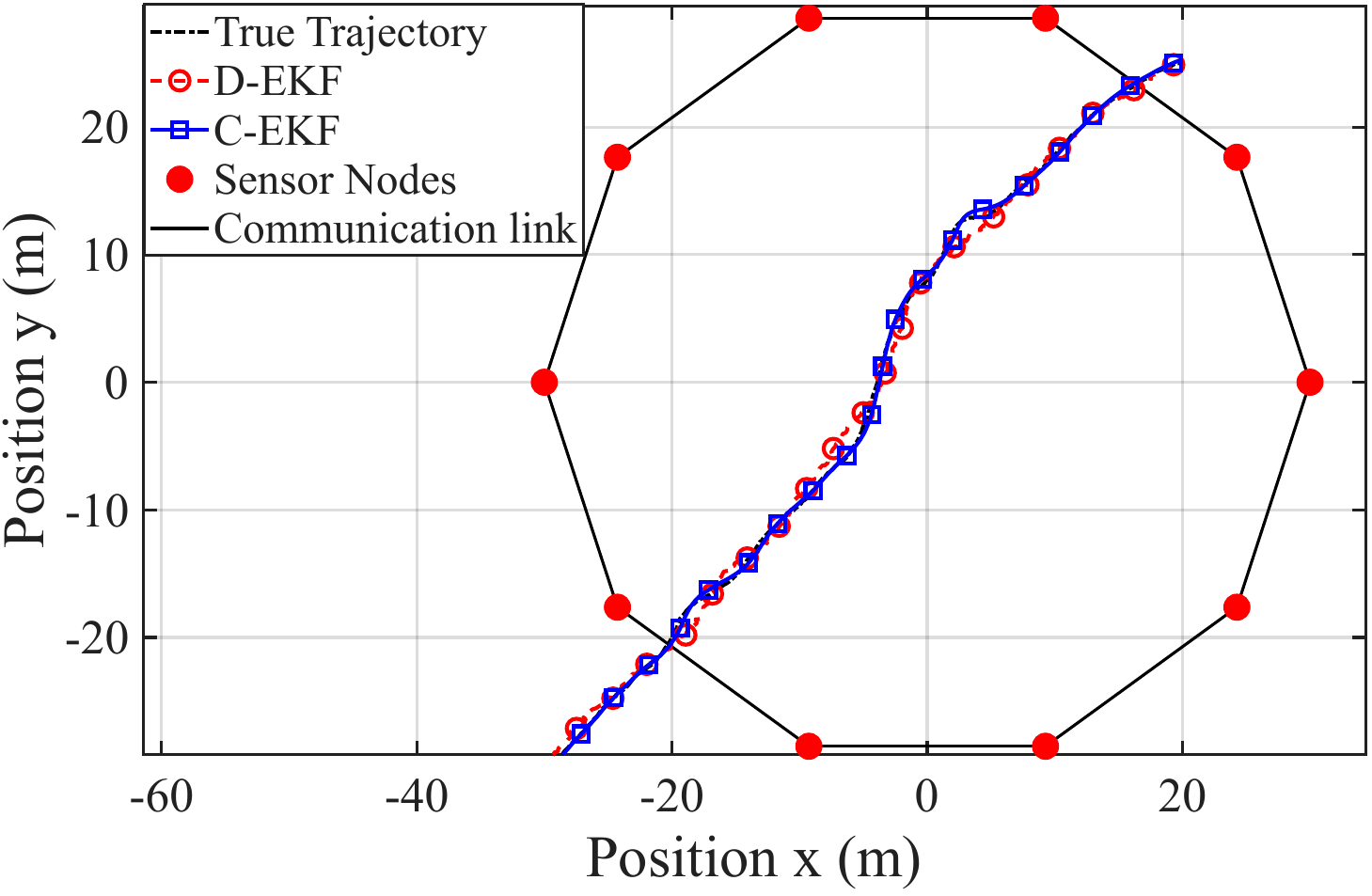}
    \caption{Tracking trajectory comparison for the proposed recursive C-EKF and D-EKF.}
    \label{fig: EKF trajectory}
\end{figure}

Fig.~\ref{fig: EKF trajectory} shows the tracking trajectory obtained by the recursive EKF approach. Both EKF estimators accurately follow the target trajectory over time, including the nonlinear portion caused by heading perturbations. Compared with the batch MAP approach, the EKF updates the state recursively as new measurements arrive and explicitly exploits the target dynamics, which enables smoother and more stable tracking. The D-EKF trajectory remains close to the C-EKF trajectory throughout the tracking interval, indicating that the proposed consensus-based distributed EKF achieves near-centralized performance using only local communication among neighboring radar nodes. The small deviations between the estimated and true trajectories are mainly observed around regions where the target heading changes, but the filter quickly corrects the estimate and maintains consistent tracking performance.
\subsection{Compare Centralized and Distributed Estimators for varying SNR}
\begin{figure*}[t]
    \centering
    \includegraphics[width=\linewidth]{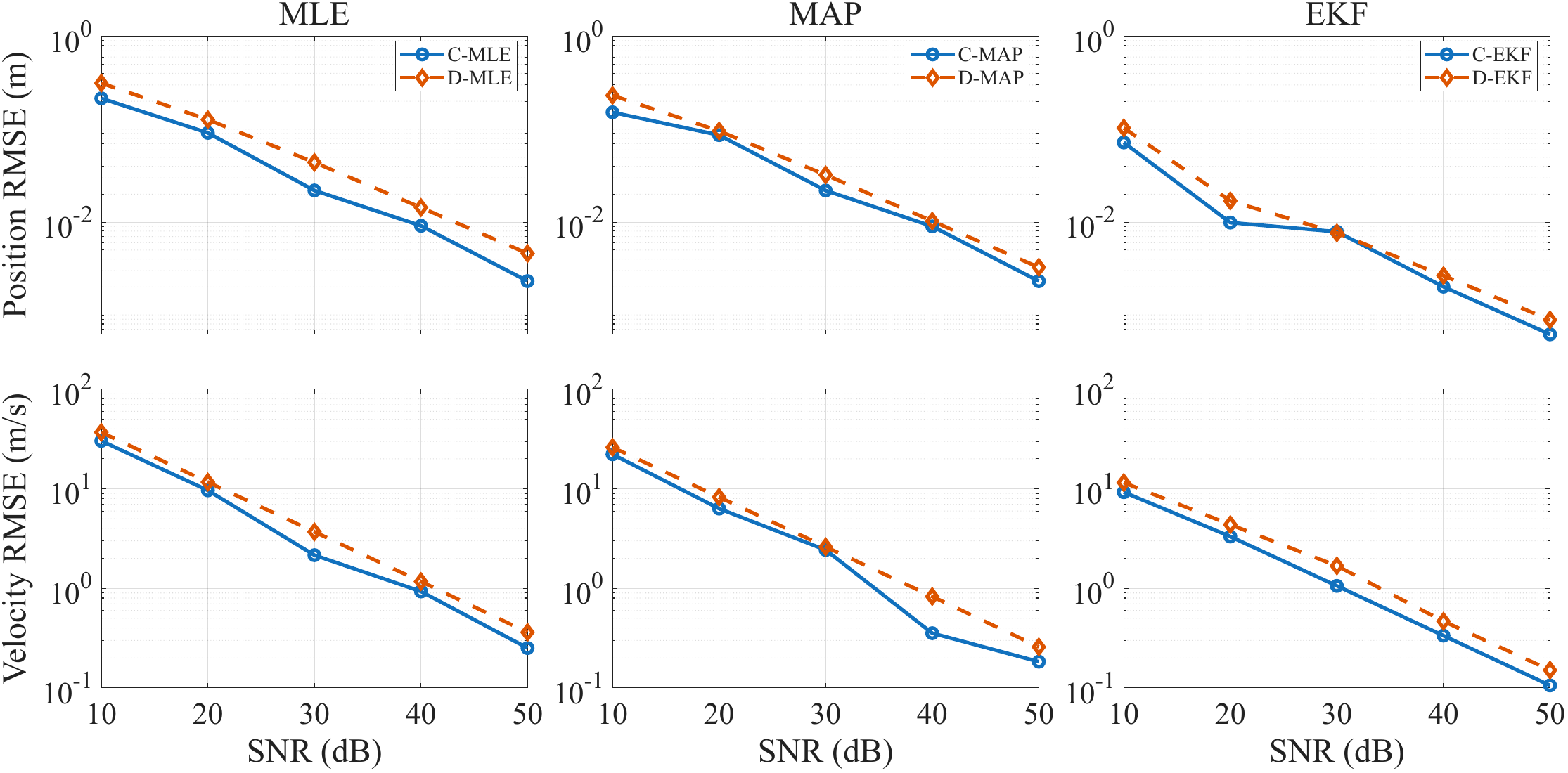}
    \caption{RMSE of estimated target states versus SNR, comparing 6 estimators: C-MLE (baseline), D-MLE (baseline), and proposed approaches: C-MAP , D-MAP , C-EKF, and D-EKF.}
    \label{fig: all estimators}
\end{figure*}

Fig.~\ref{fig: all estimators} compares the position and velocity RMSE of the target with respect to SNR for six estimators: C-MLE, D-MLE (baseline~\cite{srikar}), C-MAP, D-MAP, C-EKF, and D-EKF. Across states, the RMSE decreases monotonically as SNR increases. The RMSE on velocity is consistently higher than the position RMSE for two main reasons. First, range accuracy is primarily determined by the signal bandwidth, as shown in \eqref{eq:sigma model range}, whereas radial-velocity accuracy depends strongly on the coherent observation duration across chirps, as indicated in \eqref{eq:sigma model doppler}. Under the relatively short coherent processing interval considered in the simulations, the Doppler measurements therefore provide less accurate velocity information. Second, the nonlinear and non-convex Doppler measurement model in \eqref{eq:mea model doppler} makes the associated optimization problem more difficult to solve and may cause the consensus-based estimators to converge to local minima, further degrading velocity-estimation accuracy.

Among the batch-processing estimators, MAP consistently outperforms MLE in both centralized and distributed settings. This demonstrates the benefit of incorporating prior information from the previous tracking step. The improvement is especially visible in 10dB to 30dB SNR conditions, where the prior helps regularize the estimation problem and improves robustness against noisy measurements. The D-MAP estimator closely follows its centralized counterpart, indicating that the proposed consensus optimization formulation can achieve near-centralized performance using only local communication.

The recursive estimators (C-EKF, D-EKF) achieve the lowest RMSE among all methods. This is because the EKF explicitly exploits the target dynamics and recursively propagates both the state estimate and its uncertainty over time. Similar to the MAP case, the D-EKF remains close to the C-EKF across different SNR levels, the performance gap arises from finite consensus iterations, node-dependent linearization points, and convergence of the nonconvex distributed optimization to stationary solutions that are not necessarily globally optimal.
\subsection{Uncertainty for Recursive Estimators}
In this study, we compare the time-varying RMSE of the D-EKF estimates with the corresponding network-wide PCRLB ~\cite{PCRLB} based on the available measurements in the whole network. The time-wise RMSE are averaged over $N_\text{MC}=20$ Monte Carlo runs, $N=10$ nodes. Note that the estimation error covariance is lower bounded by the inverse of the posterior Fisher information matrix, i.e.,
$\mathrm{cov}(\hat{\bsym{\theta}}_k-\bsym{\theta}_k) \succeq \mathbf{J}_{k}^{-1}.$
For the state vector $\bsym{\theta}_k=[x_k,y_k,v_{x,k},v_{y,k}]^\top$, the covariance matrix can be written as
\begin{equation}
\mathrm{cov}(\hat{\bsym{\theta}}_k-\bsym{\theta}_k) =
    \begin{bmatrix}
        \sigma_x^2 & \ast & \ast & \ast\\
        \ast & \sigma_y^2 & \ast & \ast \\
        \ast & \ast & \sigma_{v_x}^2 & \ast\\
        \ast & \ast & \ast & \sigma_{v_y}^2
    \end{bmatrix}\text{.}
\end{equation}

We report the RMSE of the two-dimensional position and velocity and the corresponding PCRLB of the centralized estimator is also evaluated in terms of the marginal position and velocity uncertainties. Specifically, we use the diagonal entries of the covariance lower bound $\mathbf{J}k^{-1}$, which correspond to the marginal variances of $x$, $y$, $v_x$, and $v_y$. The off-diagonal entries describe cross-correlations between state components and are important for the full covariance structure. Accordingly, the root PCRLBs for position and velocity are computed as
\begin{equation}
\begin{aligned}
\mathrm{RPCRLB}{p,k}
&=
\sqrt{
[\mathbf{J}_k^{-1}]_{x,x}
+
[\mathbf{J}_k^{-1}]_{y,y}
}\text{,}\\
\mathrm{RPCRLB}_{v,k}
&=
\sqrt{
[\mathbf{J}_k^{-1}]_{v_x,v_x}
+
[\mathbf{J}_k^{-1}]_{v_y,v_y}
}\text{.}
\end{aligned}
\end{equation}

\begin{figure*}[t]
    \centering
    \includegraphics[width=\linewidth]{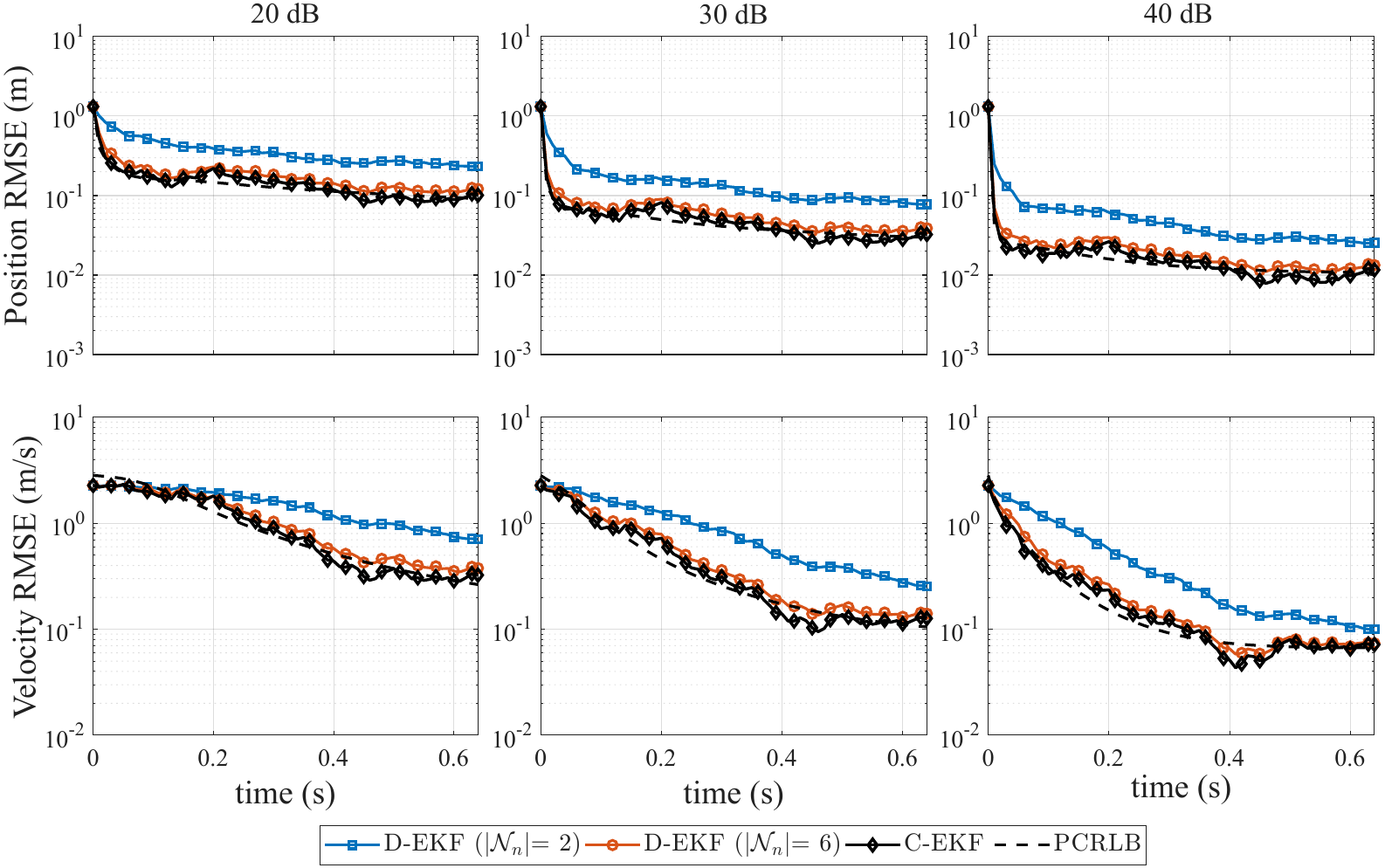}
    \caption{RMSE of the estimates w.r.t. time over different neighboring nodes. The square and circle curves correspond to the RMSE of D-EKF (proposed) with two and six cooperating radar nodes, the diamond curve correspond to RMSE of C-EKF (proposed),  dashed curve correspond to root PCRLB for position and velocity estimation.}
    \label{fig: PCRLB}
\end{figure*}

In the simulations, the Jacobian $\mbf{H}_{n,k}$ is evaluated along the ground-truth trajectory for each Monte Carlo run, and the resulting PCRLB curves are averaged over all runs. Fig.~\ref{fig: PCRLB} compares the position and velocity RMSE of the D-EKF using $|\mathcal{N}_n|=2$ and $|\mathcal{N}_n|=6$ cooperating radar nodes with the C-EKF and the PCRLB at SNR levels of 20, 30, and 40~dB. As shown in the figure, increasing the SNR consistently reduces both the position and velocity estimation errors. The C-EKF rapidly converges toward the PCRLB after the initial transient, demonstrating the benefit of jointly exploiting measurements from all radar nodes. In contrast, the D-EKF with three cooperating nodes exhibits a clear performance degradation relative to the C-EKF, highlighting the inherent trade-off between centralized and fully distributed fusion: while the centralized approach achieves near-optimal accuracy by globally fusing all measurements, the distributed implementation reduces communication and computational burden at the cost of reduced estimation performance due to limited local information exchange.

For the distributed estimators, increasing the number of cooperating nodes from two to six significantly improves the tracking accuracy. The D-EKF with six nodes consistently approaches the C-EKF more closely than the three-node configuration, since additional radar nodes provide complementary geometric and measurement information during the distributed fusion process. This improvement is particularly evident for velocity estimation, where the three-node D-EKF exhibits a larger convergence time and higher RMSE, especially at lower SNR. In contrast, the six-node D-EKF follows the centralized estimator more closely over the tracking interval. These results demonstrate that the performance of the fully distributed estimator improves as more neighboring measurements are incorporated, allowing it to progressively bridge the gap toward centralized fusion performance.

At several time instants, particularly at high SNR, the empirical C-EKF RMSE slightly falls below the PCRLB. This apparent violation is attributed to the limited number of Monte Carlo realizations ($N_{\mathrm{MC}}=20$), since the empirical RMSE is a finite-sample approximation of the expected estimation error and may fluctuate around the theoretical lower bound.

The overall results illustrate the trade-off between estimation accuracy and number of node involved in the fully distributed architecture, which removes the dependence on a centralized fusion center. The recursive computation of the  PCRLB for the D-EKF is provided in Appendix~\ref{app: FIM}.

%% file: chapter/conclusion.tex
\section{Conclusion}
\label{sec: conclusion}
% What we proposed, What we are compare with and what we achieve 
In this paper, we propose distributed MAP and distributed EKF approaches for target tracking in a radar network. Unlike traditional radar tracking systems that rely on centralized or decentralized architectures, which are vulnerable to single-point failures and require substantial communication to a fusion center, our approach enables each radar node to exchange information with its nearest neighbors and to perform estimation in a fully distributed manner. Through theoretical analysis and numerical simulations, we show that the proposed methods achieve reliable tracking performance and outperform state-of-the-art MLE-based method under a different SNR conditions.

% what's the limitation and the future work
In future work, several directions can be explored. First, the proposed framework can be extended to multi-target tracking in multi-radar networks, where data association and scalability become key challenges. Second, data-driven models can be incorporated to capture unknown or time-varying target dynamics, improving robustness when the assumed Kalman-filter dynamics are violated. Third, adaptive time-varying network topologies can be explored, since the informativeness of neighboring radar nodes depends on their relative geometry and viewing angles. Finally, alternative distributed optimization methods, such as proximal ADMM~\cite{PxADMM} and PDMM~\cite{pdmm}, can be investigated to assess potential computational and communication benefits beyond the standard consensus ADMM adopted in this study.

%% file: chapter/appendix.tex
\appendices
\section{PCRLB for Target Tracking using Radar network}
\label{app: FIM}
\setcounter{equation}{0}
\renewcommand{\theequation}{A.\arabic{equation}}
% Describe the problem brieftly again
We consider the state space model introduced in (\ref{eq: SSM}) where the target state is denoted by $\bsym{\theta}_k= [\bsym{p}_k,\bsym{v}_k]^\top=[x_k,y_k,v_{x,k},v_{y,k}]^\top$.
Let $\hat{\bsym{\theta}}_k$ denote an arbitrary estimator of the target state based on all the measurements available up to time $k$, estimated by the whole radar network. The Bayesian information inequality states that the Bayesian mean square error matrix of any estimator is lower bounded by the inverse of the posterior information matrix (PIM):
\begin{equation}
\mathbb{E}\!\left[
(\hat{\bsym{\theta}}_k-\bsym{\theta}_k)
(\hat{\bsym{\theta}}_k-\bsym{\theta}_k)^\top
\right]
\succeq
\mbf{J}_k^{-1},
\label{eq:bayesian_information_inequality}
\end{equation}
where $\mbf{J}_k$ denotes the PIM at time $k$. This bound applies to the filtering problem and does not require the estimator to be unbiased.
The PIM is propagated recursively over tracking time. The prediction step first maps the posterior information matrix of previous time step $\mbf{J}_{k-1}$ through the target dynamics. Since the dynamics model is linearized by the Jacobian $\mbf{F}_{k}$, the predicted information matrix is given by
\begin{equation}
\mbf{J}_{k|k-1}
=
\left(
\mbf{F}_{k}\mbf{J}_{k-1}^{-1}\mbf{F}_{k}^{\top}
+
\mbf{Q}_{k}
\right)^{-1}.
\label{eq:pcrlb_prediction}
\end{equation}
This equation corresponds to propagating the previous PCRLB covariance through the target dynamics and converting the result back to information form. At time $k$, each radar node contributes information through its local range and Doppler measurement model \eqref{eq: mea model}. Define
$R_{n,k}\coloneqq(x_k-x_n)^2+(y_k-y_n)^2,
\quad r_{n,k}=\sqrt{R_{n,k}}\,$,and 
$\alpha_{n,k}\coloneqq\bsym{v}_k^\top(\bsym{p}_k-\bsym{g}_n)=v_{x,k}(x_k-x_n)+v_{y,k}(y_k-y_n).$ The range derivatives are then given by 

\begin{equation} \begin{aligned} \frac{\partial r_n(\bsym{\theta}_k)}{\partial x_k} &= \frac{x_k-x_n}{\sqrt{R_{n,k}}}, \\ \frac{\partial r_n(\bsym{\theta}_k)}{\partial y_k} &= \frac{y_k-y_n}{\sqrt{R_{n,k}}}, \\ \frac{\partial r_n(\bsym{\theta}_k)}{\partial v_{x,k}} &=0, \\ \frac{\partial r_n(\bsym{\theta}_k)}{\partial v_{y,k}} &=0. \end{aligned} \label{eq:range_derivatives_short} \end{equation}

The Doppler derivatives can then be written compactly as
\begin{equation}
\begin{aligned}
\frac{\partial f_n(\bsym{\theta}_k)}{\partial x_k}
&=
-
\frac{2}{\lambda}
\frac{
v_{x,k} r_{n,k}
+
\alpha_{n,k}
\frac{x_k-x_n}{\sqrt{R_{n,k}}}
}
{R_{n,k}},
\\[2mm]
\frac{\partial f_n(\bsym{\theta}_k)}{\partial y_k}
&=
-
\frac{2}{\lambda}
\frac{
v_{y,k} r_{n,k}
+
\alpha_{n,k}
\frac{y_k-y_n}{\sqrt{R_{n,k}}}
}
{R_{n,k}},
\\[2mm]
\frac{\partial f_n(\bsym{\theta}_k)}{\partial v_{x,k}}
&=
-\frac{2}{\lambda}
\frac{x_k-x_n}
{\sqrt{R_{n,k}}},
\\[2mm]
\frac{\partial f_n(\bsym{\theta}_k)}{\partial v_{y,k}}
&=
-\frac{2}{\lambda}
\frac{y_k-y_n}
{\sqrt{R_{n,k}}}.
\end{aligned}
\label{eq:doppler_derivatives_short}
\end{equation}
Let
\begin{equation}
\begin{aligned}
\mbf{H}_{n,k}
& =
\left.
\frac{\partial \bsym{\mu}_n(\bsym{\theta})}
{\partial \bsym{\theta}}
\right|_{\bsym{\theta}=\bsym{\theta}_{n,k}} \\
& = 
\left.
\begin{bmatrix}
    \frac{\partial r_n(\bsym{\theta})}{\partial x} & \frac{\partial r_n(\bsym{\theta})}{\partial y}& 0 &0\\
    \frac{\partial f_n(\bsym{\theta})}{\partial x} & \frac{\partial f_n(\bsym{\theta})}{\partial y}& \frac{\partial f_n(\bsym{\theta})}{\partial v_x}& \frac{\partial f_n(\bsym{\theta})}{\partial v_y}
\end{bmatrix}\right|_{\bsym{\theta}=\bsym{\theta}_{n,k}}\text{,}
\label{eq:pcrlb_jacobian}
\end{aligned}
\end{equation}
denote the Jacobian of the measurement model of node $n$, which is the partial derivative of measurement model \eqref{eq: mea model} w.r.t. target state. Because the measurement model is nonlinear,
\(\mathbf{H}_{n,k}(\boldsymbol{\theta}_k)\) depends on the random
target state. Assuming conditionally independent measurement noises
with known, state-independent covariance matrices, the information
contributions from the radar nodes are additive. Therefore,
\begin{equation}
\label{eq:pcrlb_update}
\mathbf{J}_k
=
\mathbf{J}_{k|k-1}
+
\sum_{n=1}^{N}
\mathbb{E}_{\boldsymbol{\theta}_k}
\left[
\mathbf{H}_{n,k}^{\top}(\boldsymbol{\theta}_k)
\boldsymbol{\Sigma}_{n,k}^{-1}
\mathbf{H}_{n,k}(\boldsymbol{\theta}_k)
\right].
\end{equation}
In the simulations, the expectation in
\eqref{eq:pcrlb_update} is approximated by averaging the measurement
information over \(N_{\text{MC}}\) Monte Carlo trajectories:
\begin{equation}
\widehat{\mathbf{I}}_{k}^{\text{meas}}
=
\frac{1}{N_{\text{MC}}}
\sum_{q=1}^{N_{\text{MC}}}
\sum_{n=1}^{N}
\mathbf{H}_{n,k}^{\top}
\left(\boldsymbol{\theta}_{k}^{(q)}\right)
\boldsymbol{\Sigma}_{n,k}^{-1}
\mathbf{H}_{n,k}
\left(\boldsymbol{\theta}_{k}^{(q)}\right).
\end{equation}
The posterior information matrix is then approximated as
\begin{equation}
\mathbf{J}_k
=
\mathbf{J}_{k|k-1}
+
\widehat{\mathbf{I}}_{k}^{\mathrm{meas}},
\end{equation}
and the corresponding posterior covariance lower bound is
\(\mathbf{J}_k^{-1}\). For each state component $d \in \{x,y,v_x,v_y\}$, the root PCRLB used in the numerical results is computed as
\begin{equation}
\mathrm{RPCRLB}_{d,k}
=
\sqrt{
\left[
\mbf{J}_{k}^{-1}
\right]_{d,d}
}.
\label{eq:rpcrlb}
\end{equation}
The recursion is initialized from the prior covariance as
\begin{equation}
\mbf{J}_0 = \mbf{P}_0^{-1}.
\label{eq:pcrlb_initialization}
\end{equation}